\documentclass[aps,PRB,reprint,superscriptaddress,preprintnumbers]{revtex4-2}
\usepackage[T1]{fontenc}
\usepackage{times}
\usepackage{graphicx,color}
\usepackage{amsfonts,amsmath,amssymb,amsbsy}
\usepackage[colorlinks=true, linkcolor=blue, citecolor=blue, urlcolor=blue]{hyperref}
\usepackage{float}

\def\D{{\boldsymbol{\mathcal{D}}}}

\def\I{{\boldsymbol{\mathcal{I}}}}
\def\m{\boldsymbol{m}}
\def\n{\boldsymbol{n}}
\def\s{\boldsymbol{s}}
\def\He{\boldsymbol{H}}
\def\f{\boldsymbol{f}}
\def\p{\boldsymbol{p}}

\def\T{\boldsymbol{T}}
\def\G{\boldsymbol{G}}
\def\R{\boldsymbol{R}}
\def\0{\boldsymbol{0}}
\def\sin{\text{sin}}
\def\cos{\text{cos}}
\let\mathbf=\boldsymbol

\def\emph#1{\textcolor{blue}{#1}}
\begin{document}

\title{Dynamics of ferrimagnetic skyrmionium driven by spin-orbit torque}

\author{Xue Liang}
\affiliation{School of Science and Engineering, The Chinese University of Hong Kong, Shenzhen, Guangdong 518172, China}

\author{Xichao Zhang}
\affiliation{Department of Electrical and Computer Engineering, Shinshu University, Wakasato 4-17-1, Nagano 380-8553, Japan}

\author{Laichuan Shen}
\affiliation{School of Science and Engineering, The Chinese University of Hong Kong, Shenzhen, Guangdong 518172, China}

\author{Jing Xia}
\affiliation{College of Physics and Electronic Engineering, Sichuan Normal University, Chengdu 610068, China}

\author{Motohiko Ezawa}
\email[Email:~]{ezawa@ap.t.u-tokyo.ac.jp}
\affiliation{Department of Applied Physics, The University of Tokyo, 7-3-1 Hongo, Tokyo 113-8656, Japan}

\author{\\ Xiaoxi Liu}
\affiliation{Department of Electrical and Computer Engineering, Shinshu University, Wakasato 4-17-1, Nagano 380-8553, Japan}

\author{Yan Zhou}
\email[Email:~]{zhouyan@cuhk.edu.cn}
\affiliation{School of Science and Engineering, The Chinese University of Hong Kong, Shenzhen, Guangdong 518172, China}

\begin{abstract}
Magnetic skyrmionium is a skyrmion-like spin texture with nanoscale size and high mobility.
It is a topologically trivial but dynamically stable structure, which can be used as a non-volatile information carrier for next-generation spintronic storage and computing devices.
Here, we study the dynamics of a skyrmionium driven by the spin torque in a ferrimagnetic nanotrack.
It is found that the direction of motion is jointly determined by the internal configuration of a skyrmionium and the spin polarization vector. Besides, the deformation of a skyrmionium induced by the intrinsic skyrmion Hall effect depends on both the magnitude of the driving force and the net angular momentum.
The ferrimagnetic skyrmionium is most robust at the angular momentum compensation point, whose dynamics is quite similar to the skyrmionium in antiferromagnet.
The skyrmion Hall effect is perfectly prohibited, where it is possible to observe the position of the skyrmionium by measuring the magnetization.
Furthermore, the current-induced dynamics of a ferrimagnetic skyrmionium is compared with that of a ferromagnetic and antiferromagnetic skyrmionium. We also make a comparison between the motion of a ferrimagnetic skyrmionium and a skyrmion.
Our results will open a new field of ferrimagnetic skyrmioniums for future development of ferrimagnetic spintronics devices.
\end{abstract}

\date{\today}

\maketitle

\section{Introduction}
\label{se:Introduction}

Spintronics is an important field, which has led to many information processing applications, including memory and logic computing devices~\cite{Torrejon_NAT2017,Parkin_NATTECH2015,Fert_NATTECH2013,Parkin_SCI2008,Finocchio_JDA2016,Iwasaki_NATTECH2013,Sampaio_NATTECH2013,Tomasello_SCIREP2014,Zhouy_NSR2019,Xichao_SCIREP9400,Luo_NANO2018,Xing_PRB2016}.
The primary tasks in the field of spintronics include the understanding of the fundamental physical phenomena associated with magnetic spins and the exploration of various effective methods to control spins~\cite{Zutic_RMP2004,Jungwirth_NATTECH2016}.
Ferromagnetic (FM) materials are vital for spin-texture-based devices~\cite{Torrejon_NAT2017,Parkin_NATTECH2015,Fert_NATTECH2013,Parkin_SCI2008,Finocchio_JDA2016,Zhouy_NSR2019,Xing_PRB2016,Zutic_RMP2004}; however, the scaling limitation as well as the limited operation speed may hinder the use of ferromagnets in future spintronic devices~\cite{Jungwirth_NATTECH2016,Yang_NATTECH2015,Baltz_RMP2018,Buttner_SCIREP2018}.
Therefore, it is necessary to find some different types of magnetic materials to host smaller spin textures and to realize faster spin dynamics.

Antiferromagnets are predicted theoretically to stabilize nanoscale spin textures showing ultra-fast spin dynamics, which have been an emerging topic in recent years~\cite{Jungwirth_NATTECH2016,Baltz_RMP2018,Barker_PRL2016,Shiino_PRL2016,Gomonay_PRL2016,Xichao_SCIREP24795}.
However, the direct manipulation and detection of antiferromagnetic (AFM) spin textures are still challenges in experiments due to the fact that a perfect antiferromagnet shows zero net magnetic moment.
Hence, recently much attention has being paid to a special ordered spin system, that is, the ferrimagnetic (FiM) system, where the magnetic moments from two inequivalent sublattices are coupled in an AFM configuration. This system holds desirable qualities of both ferromagnets and antiferromagnets, such as the measurable net magnetization and the fast dynamics of spins, which may provide a promising material platform for the research of spintronics devices.

One class of ferrimagnets is the rare earth (RE)-transition metal (TM) compounds~\cite{Stanciu_PRB2006,Kim_NATMAT2017,Woo_NATCOM2018,Caretta_NATTECH2018,Binder_PRB2006,Kim_PRB2017}, in which the RE and TM elements have different Land{\'e} $g$ factors, corresponding to the gyromagnetic ratio $\gamma$. Consequently, FiM systems have two significant compensation points, that is, the magnetization compensation point and the angular momentum compensation point. In particular, at the angular momentum compensation point, the AFM dynamics can be achieved due to the complete cancellation of angular momentum, and the resultant magnetization is nonzero, which allows the manipulation of ferrimagnets by an external field similar to the case of ferromagnets~\cite{SHOh_PRB2017,SHOh_PRB2019}. These remarkable features enable ferrimagnets to solve the difficulty on the detection of antiferromagnets and overcome the low-speed limitation of ferromagnets, thus being a candidate system for studying spin structures and dynamics.

On the other hand, future spintronic devices are expected to store and process digital information with an ultra-fast operation speed and ultra-high storage density.
The topologically protected spin texture called the skyrmion is a promising building block for future spintronic devices~\cite{Fert_NATTECH2013,Finocchio_JDA2016,Iwasaki_NATTECH2013,Sampaio_NATTECH2013,Zhouy_NSR2019,Muhlbauer_SCI2009,Yu_NAT2010,Nagaosa_NATTECH2013,Xichao_JPC2020}, which can be used to encode bits in racetrack-based devices and has been extensively studied in both theoretical and experimental approaches~\cite{Finocchio_JDA2016,Fert_NATTECH2013,Tomasello_SCIREP2014,Xichao_SCIREP9400,Luo_NANO2018,Xing_PRB2016}.
The magnetic skyrmion has some unique properties, such as the nanoscale size and low depinning current density, which provide a certain possibility to build spintronic devices with high performance.
However, skyrmions may deviate from the direction of the driving force induced by external stimulus, such as the spin-polarized current, since the topology-dependent Magnus force introduces a transverse velocity, which is known as the skyrmion Hall effect~\cite{Jiang_NATPHY2016,Litzius_NATPHY2016,Zang_PRL2011}.
Such a phenomenon will lead to a serious accumulation or annihilation of skyrmions at the boundary of devices. To avoid this detrimental effect, tremendous efforts have been devoted to the realization of the in-line motion of skyrmions by using various methods, including the modification of magnetic properties~\cite{Lai_SCIREP2017,Kentaro_NANO2021,Romeo_NANO2021}, the use of synthetic AFM skyrmions~\cite{Xichao_NC2016,Dohi_NC2019,Legrand_MAT2020,ZhangX_PRB2016}, and the AFM skyrmions~\cite{Barker_PRL2016,Xichao_SCIREP24795}.

\begin{figure}[h]
\centerline{\includegraphics[width=0.53\textwidth]{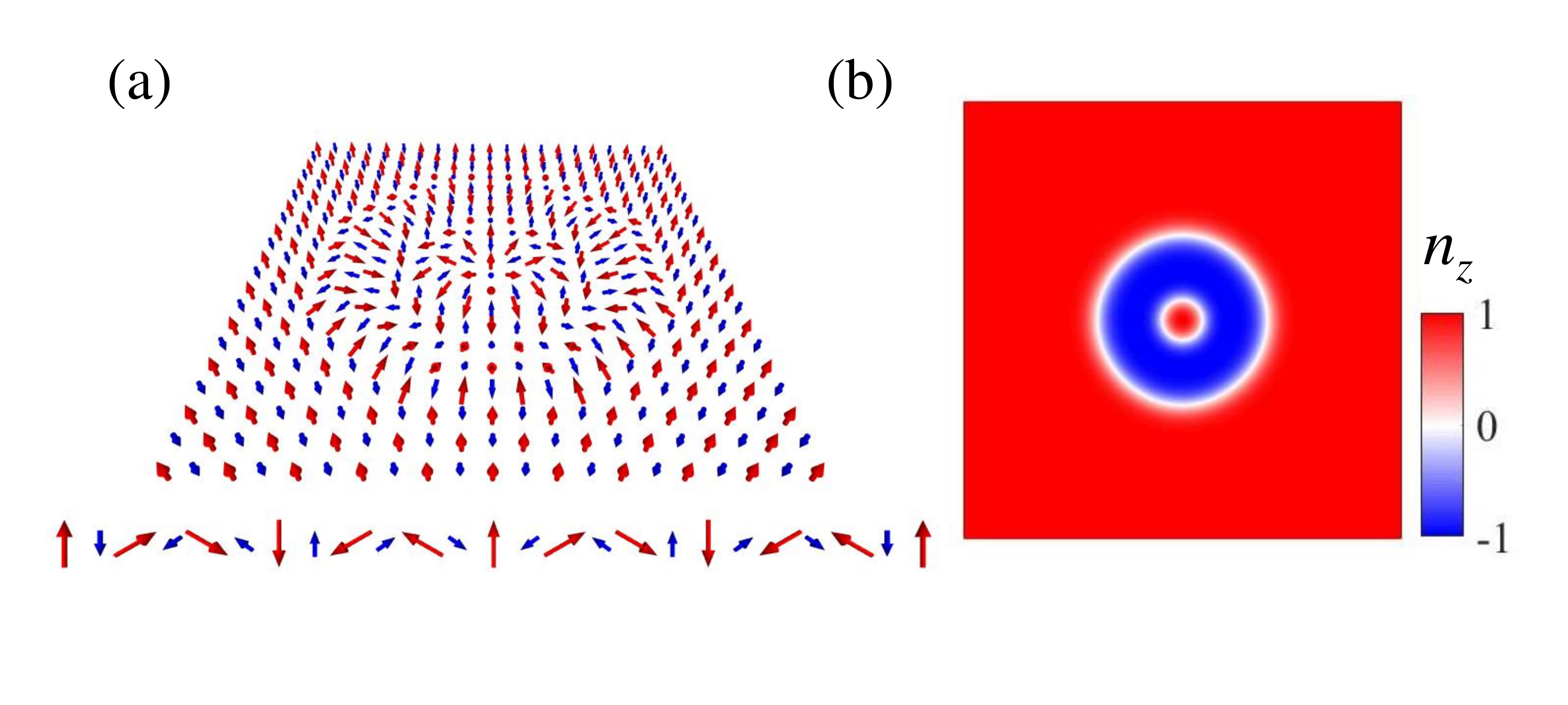}}
\caption{%
Schematic diagram of an isolated ferrimagnetic skyrmionium in a two-dimensional film.
(a) Illustration of a N{\'e}el-type ferrimagnetic skyrmionium spin structure with the two anti-parallel magnetic sublattices. The magnetization orientation along any centerline through the skyrmionium core is sketched below. Red and blue colors indicate the sublattices 1 and 2, respectively.
(b) The simulation snapshot of a relaxed ferrimagnetic skyrmionium, where the $z$-component of N{\'e}el vector $\boldsymbol{n}$ is coded in the pixel color as shown in the color bar on the right. 
}
\label{FIG1}
\end{figure}

Similar to a synthetic AFM bilayer skyrmion, the skyrmionium can be regarded as a combination of two skyrmions with topological charges $Q=+1$ and $Q=-1$. Since the magnitude of Magnus force is proportional to the topological charge and its direction also depends on the sign of $Q$, these two opposite Magnus forces acting on the skyrmionium can be compensated completely, eliminating the skyrmion Hall effect.~\cite{Xichao_PRB2016,Obadero_PRB2020}.
Several works have demonstrated that the zero net topological charge of a skyrmionium allows its motion along the driving current without transverse drift~\cite{Xichao_PRB2016,Obadero_PRB2020,olesnikov_SCIREP2018,LISAI_APL2018,WangJ_APL2020,Gobel_SCIREP2019,Laichuan_PRA2019}.
Most recently, such a spin texture has been experimentally observed in FiM multilayers~\cite{Seng_arX2021} and its topological counterpart, which is called the bimeronium, has also been studied in frustrated magnets~\cite{Xichao_APL2021}.
However, the dynamics of a FiM skyrmionium induced by external stimulus still remain elusive.

In this work, we systematically investigate the spin current-induced dynamics of a skyrmionium in a FiM nanotrack by using both theoretical and numerical methods. It is found that the FiM skyrmionium can reach a higher speed without showing a severe distortion compared with the FM skyrmionium. The dynamics of a FiM skyrmionium at the compensation point of angular momentum is comparable with the case of an AFM skyrmionium because of the vanishing angular momentum. Another prominent feature is that it is possible to observe the position of the skyrmionium due to a nonzero magnetization at this point.
%
 
\section{Model and Methods}
\label{se:Model and Methods}

We consider a basic RE-TM FiM film consisting of two sublattices, where the unit magnetic moments are $\boldsymbol{s}_{1}(\boldsymbol{r},t)$ and $\boldsymbol{s}_{2}(\boldsymbol{r},t)$.
The N{\'e}el vector and total unit magnetization are defined as $\boldsymbol{n}(\boldsymbol{r},t)=(\boldsymbol{s}_{1}-\boldsymbol{s}_{2})/{2}$ and $\boldsymbol{m}(\boldsymbol{r},t)=(\boldsymbol{s}_{1}+\boldsymbol{s}_{2})/{2}$, respectively.
In the continuum approximation, the energy functional is given by ~\cite{Tveten_PRB2016}
\begin{equation}
\begin{split}
\label{eq:E}
E_{\text{total}}=&\int \{\frac{\lambda}{2}\boldsymbol{m}^2+\frac{A}{2}[(\boldsymbol{{\nabla}n})^2+\partial_{x}\boldsymbol{n}\cdot\partial_{y}\boldsymbol{n}]\\
&+L\boldsymbol{m}\cdot(\partial_{x}\boldsymbol{n}+\partial_{y}\boldsymbol{n})-\frac{K}{2}{n^2_{z}}+\varepsilon_{\text{DMI}}\} {d}V,
\end{split}
\end{equation}
where $\lambda$, $A$ and $L$ are the homogeneous, inhomogeneous and parity-breaking exchange constants, respectively~\cite{Shiino_PRL2016,Tveten_PRB2016}. $K$ is the perpendicular magnetic anisotropy (PMA) constant.
The last term $\varepsilon_{\text{DMI}}=(D/2)[n_{z}(\boldsymbol{\nabla}\cdot\boldsymbol{n})-(\boldsymbol{n}\cdot\boldsymbol{\nabla})n_{z}]$ is the interface-induced Dzyaloshinskii-Moriya interaction (DMI) energy density~\cite{Rohart_PRB2013,Dzyaloshinsky_PCS1958,Moriya_PR1960}, which stabilizes the N{\'e}el-type skyrmionium, as illustrated in Fig.~\ref{FIG1}.
By taking the functional derivative of the total energy density $\varepsilon$, we obtain the effective fields $\boldsymbol{f}_{\boldsymbol{n}}=-{\delta\varepsilon}/({\mu_{0}\delta}\boldsymbol{n})$ and $\boldsymbol{f}_{\boldsymbol{m}}=-{\delta\varepsilon}/({\mu_{0}\delta}\boldsymbol{m})$ for the N{\'e}el vector and total unit magnetization, respectively.

For the spin dynamics driven by the spin-orbit torque (SOT), the following coupled equations of motion are constructed (see Appendix~\ref{se:Appendix A} for more details on the derivation)
\begin{eqnarray}
\rho\dot{\n}&=&\f_{\boldsymbol{m}}\times\boldsymbol{n}+\alpha(\rho\boldsymbol{n}\times\dot{\boldsymbol{m}}+\sigma\boldsymbol{n}\times\dot{\boldsymbol{n}}) \nonumber\\
&&+u_{2}\boldsymbol{n}\times(\boldsymbol{p}\times\boldsymbol{n})+\boldsymbol{T}^{n}_{\text{nl}}, \label{eq:LLGn} \\ 
\rho\dot{\boldsymbol{m}}+\sigma\dot{\boldsymbol{n}}&=&\boldsymbol{f}_{\boldsymbol{m}}\times\boldsymbol{m}+\boldsymbol{f}_{\boldsymbol{n}}\times\boldsymbol{n}+\alpha\rho\boldsymbol{n}\times\dot{\boldsymbol{n}} \nonumber \\
&&+u_{1}\boldsymbol{n}\times(\boldsymbol{p}\times\boldsymbol{n})+\boldsymbol{T}^{m}_{\text{nl}},\label{eq:LLGm} 
\end{eqnarray}
where $\rho=M_{1}/\gamma_{1}+M_{2}/\gamma_{2}$, $\sigma=M_{1}/\gamma_{1}-M_{2}/\gamma_{2}$ is the spin density that is used to quantify the net angular momentum, e.g., $\sigma=0$ denotes the compensation point of angular momentum. $\alpha$ is the damping coefficient, $\boldsymbol{p}=(p_x,p_y,0)$ is the polarization vector of the spin current, $u_{1}=\beta_{1}+\beta_{2}$ and $u_{2}=\beta_{1}-\beta_{2}$ are the parameters of SOT with $\beta_{i}=\mu_{\text{B}}\theta_{\text{SH}}j/(\gamma_{i}et_{\text{z}})$. $M_{i}$ and $\gamma_{i}$ are the saturation magnetization and the gyromagnetic ratio for the sublattice $i$, $\mu_{\text{B}}$ is the Bohr magneton, $j$ is the driving current density, $\theta_{\text{SH}}$ is the spin Hall angle fixed at $0.1$ in this work, $e$ is the electron charge and $t_{z}=1.0$ nm is the thickness of the FiM film. Considering the fact that $|\boldsymbol{m}|\ll|\boldsymbol{n}|\approx 1$ for the colinear ferrimagnets, $\boldsymbol{T}^{n}_{\text{nl}}=u_{1}\boldsymbol{n}\times(\boldsymbol{p}\times\boldsymbol{m})$ and $\boldsymbol{T}^{m}_{\text{nl}}=\alpha\sigma\boldsymbol{n}\times\dot{\boldsymbol{m}}+u_{2}\boldsymbol{n}\times(\boldsymbol{p}\times\boldsymbol{m})$ are the weak nonlinear terms that will be discarded in the following derivation.

Substituting $\boldsymbol{f}_{\boldsymbol{m}}=-[\lambda\boldsymbol{m}+L(\partial_{x}\boldsymbol{n}+\partial_{y}\boldsymbol{n})]/\mu_{0}$ into Eq.~\ref{eq:LLGn} and combining with Eq.~\ref{eq:LLGm}, we obtain the closed equation in terms of the N{\'e}el vector $\boldsymbol{n}$, given as
\begin{equation}
\label{eq:Closen} 
\frac{\mu_{0}\rho^2}{\lambda}\boldsymbol{n}\times(\boldsymbol{n}\times\ddot{\boldsymbol{n}})=\sigma\boldsymbol{n}\times\dot{\boldsymbol{n}}+\alpha\rho\dot{\boldsymbol{n}}+u_{1}\boldsymbol{p}\times\boldsymbol{n}-\boldsymbol{T}_0.
\end{equation}
Here, $(1+\alpha^2)\approx 1$, $\lambda=4A/a^2$ and $L=\sqrt{2}A/a$ are used with $a$ being the lattice constant, $\boldsymbol{T}_0=(\mu_{0}\rho{u_2}/\lambda)\boldsymbol{n}\times(\boldsymbol{p}\times\dot{\boldsymbol{n}})+\boldsymbol{n}\times(\boldsymbol{f}^\ast_{n}\times\boldsymbol{n})-({Lu_{2}}/{\lambda})\boldsymbol{n}\times\{[\boldsymbol{p}\times(\partial_{x}+\partial_{y})\boldsymbol{n}]\times\boldsymbol{n}\}$, $\boldsymbol{f}^\ast_{n}=\{{A^\ast}\Delta\boldsymbol{n}+Kn_{z}\boldsymbol{e}_{z}+D[\partial_{x}n_{z}\boldsymbol{e}_{x}+\partial_{y}n_{z}\boldsymbol{e}_{y}-(\partial_{x}n_{x}+\partial_{y}n_{y})\boldsymbol{e}_{z}]\}/\mu_{0}$ is the reduced effective field with $A^\ast=A/2$~\cite{Laichuan_PRA2019}.

In order to derive the analytical velocity from the motion Eqs.~\ref{eq:LLGn} and ~\ref{eq:LLGm}, we also use the collective coordinate approach in FiM as proposed in the FM system~\cite{Thiele_PRL1973}, with which the magnetic soliton is regarded as a rigid body and the time-dependent order parameter $\boldsymbol{n}$ need to be rewritten by the soliton position $\boldsymbol{R}(t)$, $\boldsymbol{n}(\boldsymbol{r},t)=\boldsymbol{n}[\boldsymbol{r}-\boldsymbol{R}(t)]$.
Note that the time derivatives of N{\'e}el vector $\boldsymbol{n}$ is $\dot{\boldsymbol{n}}=-\partial_{i}\boldsymbol{n}\dot{R_{i}}$, and then $\ddot{\boldsymbol{n}}$ becomes $\ddot{\boldsymbol{n}}=-\partial_{i}\boldsymbol{n}\ddot{R_{i}}$, discarding the quadratic term $\partial_{ii}\boldsymbol{n}{\dot{R^2_{i}}}$. After taking the scalar product of Eq.~\ref{eq:Closen} with $\partial_{i}\boldsymbol{n}$, and integrating over the space, the motion equation of a centrosymmetric magnetic soliton is given by
\begin{equation}
\begin{split}
\label{eq:Thiele} 
M\ddot{\boldsymbol{R}}+\sigma\boldsymbol{G}\times\dot{\boldsymbol{R}}+\alpha\rho\boldsymbol{\mathcal{D}}\dot{\boldsymbol{R}}-u_{1}\boldsymbol{\mathcal{I}}\boldsymbol{p}=\boldsymbol{0},
\end{split}
\end{equation}
where $M=\mu_{0}\rho^2d/\lambda$ is the effective mass of the magnetic soliton with $d=d_{ii}=\int \partial_{i}\boldsymbol{n}\cdot{\partial_{i}\boldsymbol{n}} dxdy$, $\boldsymbol{G}=4{\pi}Q\boldsymbol{e}_{z}$ is the gyrovector with the topological charge $Q$ defined as $Q=(1/4\pi)\int \boldsymbol{n}\cdot(\partial_{x}\boldsymbol{n}\times\partial_{y}\boldsymbol{n}) dxdy$. The third and the fourth terms are the dissipative force due to the nonzero Gilbert damping and the driving force induced by SOT, respectively, where the components of tensors $\D$ and $\I$ are given by $D_{ij}=\delta_{ij}d$ and $I_{ij}=\int (\boldsymbol{n}\times\partial_{i}\boldsymbol{n})_{j} dxdy$.
The axisymmetric spin configurations are normally described by $\boldsymbol{n}=[$sin$\theta(r)$cos$\Phi(\varphi),\;$sin$\theta(r)$sin$\Phi(\varphi),\;$cos$\theta(r)]$, where $(r,\varphi)$ are the polar coordinates. Consequently, the gyrovector $\boldsymbol{G}$ (or the topological charge $Q$) and the dissipative tensor $\boldsymbol{\mathcal{D}}$ are determined by the angle $\theta$ that is the angle between the N{\'e}el vector $\boldsymbol{n}$ and the $z$-axis, while the tensor $\boldsymbol{\mathcal{I}}$ depends on both $\theta(r)$ and $\Phi(\varphi)$, as shown in Appendix~\ref{se:Appendix A}.
For the steady motion, $\ddot{\boldsymbol{R}}$ should be zero, so that we obtain the velocity $\boldsymbol{v}=\dot{\boldsymbol{R}}$ of the soliton in the following matrix  
\begin{equation}
\label{eq:v} 
\begin{bmatrix}
v_{x}\\
v_{y}
\end{bmatrix}
=\frac{u_{1}}{{\alpha}^2{\rho}^2d^2+{\sigma}^2G^2}
\begin{bmatrix}
{\alpha}{\rho}d & {\sigma}G \\
{-\sigma}G & {\alpha}{\rho}d
\end{bmatrix}\begin{bmatrix}
I_{xx} & I_{xy} \\
I_{yx} & I_{yy}
\end{bmatrix}\begin{bmatrix}
p_{x}\\
p_{y}
\end{bmatrix}.
\end{equation}
The full derivation can be found in Appendix~\ref{se:Appendix A}.

In this work, we focus on the FiM skyrmionium as illustrated in Fig.~\ref{FIG1}, where the N{\'e}el vector $\boldsymbol{n}$ rotates $2\pi$ from the center to the edge of the skyrmionium (i.e., $\theta(0)=0$ and $\theta(\infty)=2\pi$), where the topological charge $Q=0$.
Therefore, for a N{\'e}el-type skyrmionium, $\Phi=\varphi$, $I_{xy}=-I_{yx}=I=\pi\int (r\theta_{r}+$sin$\theta$cos$\theta)dr$ and $I_{xx}=I_{yy}=0$, the analytical steady velocity of the FiM skyrmionium is simplified to
\begin{equation}
\label{eq:TvN} 
\left[
\begin{array}{c}
v_{x}\\
v_{y}
\end{array}
\right]_{\text{N{\'e}el}}=\frac{u_{1}I}{\alpha\rho d}\left[
\begin{array}{c}
p_{y}\\
-p_{x}
\end{array}
\right].
\end{equation}
Similarly, for the Bloch-type skyrmionium stabilized by the bulk DMI $\varepsilon_{\text{DMI}}=(D/2)\boldsymbol{n}\cdot({\nabla}\times\boldsymbol{n})$, $\Phi=\varphi+\pi/2$, $I_{xx}=I_{yy}=-I$ and $I_{xy}=I_{yx}=0$, the velocity is given by
\begin{equation}
\label{eq:TvB} 
\left[
\begin{array}{c}
v_{x}\\
v_{y}
\end{array}
\right]_{\text{Bloch}}=\frac{-u_{1}I}{\alpha\rho d}\left[
\begin{array}{c}
p_{x}\\
p_{y}
\end{array}
\right].
\end{equation}

To confirm the analytical prediction on the skyrmionium velocity, we numerically solve the motion Eqs.~\ref{eq:LLGn} and ~\ref{eq:LLGm} by using the method shown in the Ref.~\onlinecite{Serpico_JAP2001}. In addition, the position of the FiM skyrmionium is defined as follow ~\cite{Laichuan_PRA2019,Komineas_PRB2015,Komineas_PRB20152}
\begin{equation}
\begin{split}
\label{eq:Ri} 
R_{i}=\frac{\int i(1-n_{z}) dxdy}{\int (1-n_{z}) dxdy},  \qquad  i=x,y. 
\end{split}
\end{equation}
We can therefore get the position of the FiM skyrmionium evolving in time, and then obtain the  velocity $(v_{x}, v_{y})=(\dot{R}_x, \dot{R}_y)$ numerically. According to the above analytical derivation, the dynamics of the spins in a ferrimagnetic system are independent of the specific magnetization on two sublattices, but are strongly related to $\rho$ and the total spin density $\sigma$. For simplicity, we assume that the magnetic moment $M_{1}$ is fixed at 710 kA$\cdot$m$^{-1}$, and the gyromagnetic ratio $\gamma_{1}=2.211{\times}10^5$m${\cdot}$A$^{-1}{\cdot}$s$^{-1}$ and $\gamma_{2}=1.1\gamma_{1}$~\cite{Kim_NATMAT2017,SHOh_PRB2017,Kittel_PRB1949,Hirata_PRB2018}.
Therefore, we change the ratio $\eta=M_{2}/M_{1}$ between the magnetic moments $M_{1}$ and $M_{2}$ to modify the spin density $\sigma$ (or $\rho$). Other magnetic parameters are adopted from Ref.~\onlinecite{SHOh_PRB2017,Kim_NATMAT2017}: $A=11.5$ pJ$\cdot$m$^{-1}$, $D=1.85$ mJ$\cdot$m$^{-2}$, $K=0.438$ MJ$\cdot$m$^{-3}$, $\lambda=263.3$ MJ$\cdot$m$^{-3}$, $L=38.91$ mJ$\cdot$m$^{-2}$, $\alpha=0.1\sim0.3$ with a default value of 0.2. The mesh size of $1\times1\times1$ nm$^3$ is used to discretize the FiM nanotrack with $l=500$ nm and $w=150$ nm.
%

\section{Results and Discussion}
\label{se:Results and Discussion}

\begin{figure}[h!]
\centerline{\includegraphics[width=0.49\textwidth]{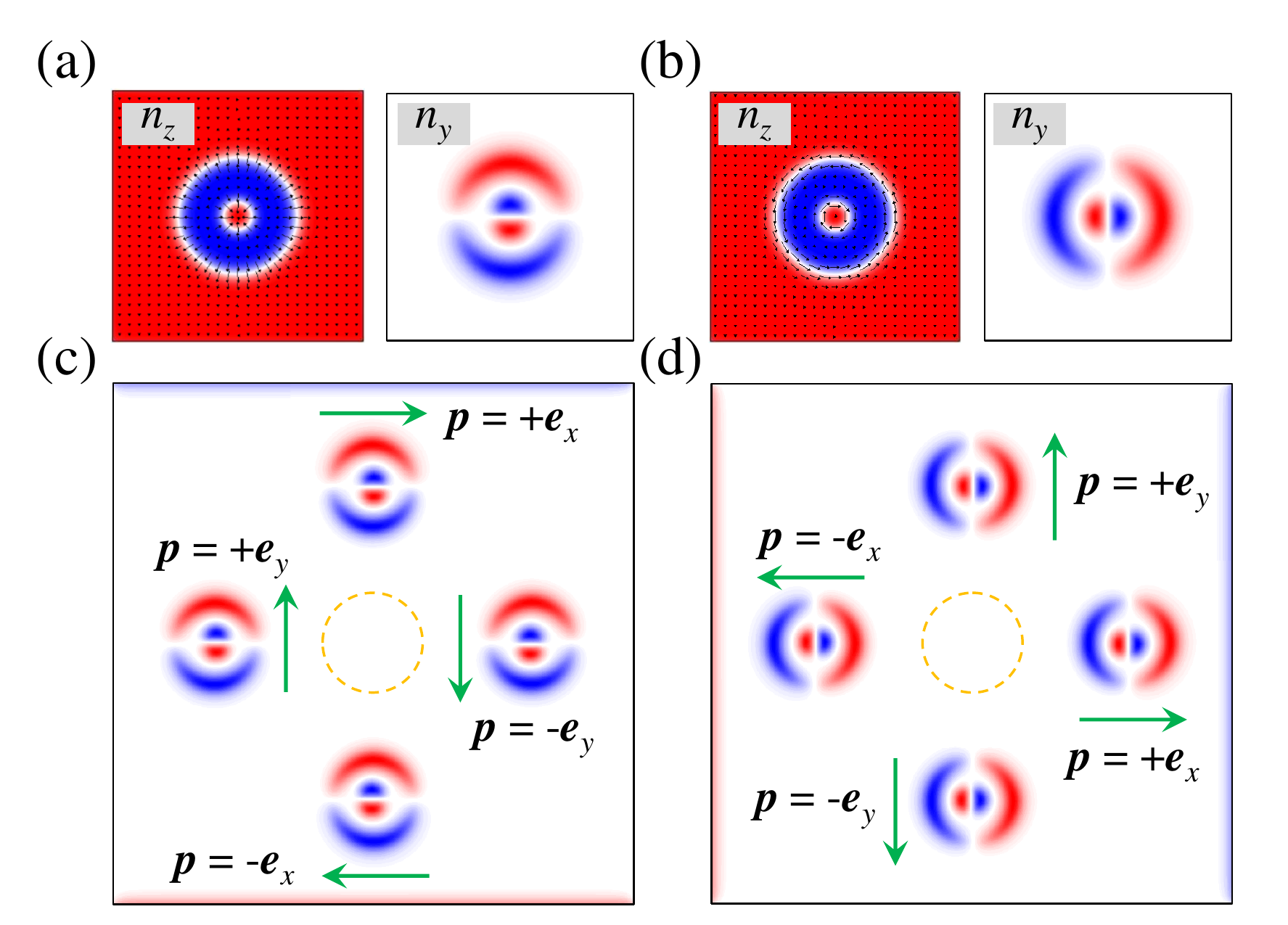}}
\caption{%
Comparison between the motion of a N{\'e}el-type skyrmionium and a Bloch-type skyrmionium.
The top-views of the (a) N{\'e}el-type skyrmionium and the (b) Bloch-type skyrmionium with the in-plane component of the order parameter $\boldsymbol{n}$ indicated by the black arrows (left), and the corresponding $y$-component of these skyrmioniums (right).
(c-d) The motion of these two types of skyrmioniums driven by the spin current with different $\boldsymbol{p}$, where the yellow dashed circle represents the initial position of the skyrmionium and green arrows denote the polarization direction.
}
\label{FIG2}
\end{figure}

From the equations (~\ref{eq:TvN}) and (~\ref{eq:TvB}), one can find that the velocity of a FiM skyrmionium depends on both the internal spin configuration (i.e., the N{\'e}el-type and the Bloch-type) and the polarization direction $\boldsymbol{p}$. Assuming that the spin current is generated from the heavy metal by the spin Hall effect, the polarization vector is thus given by $\boldsymbol{p}=\boldsymbol{e}_{z}\times\boldsymbol{j}_{e}$, where $\boldsymbol{e}_{z}$ is the unit vector normal to the film and $\boldsymbol{j}_{e}$ is the direction of the electron flow. As shown in Fig.~\ref{FIG2}, the N{\'e}el-type FiM skyrmionium moves perpendicular to the polarization direction (or parallel to the driving current) without showing the skyrmion Hall effect. However, compared with the N{\'e}el-type skyrmionium, the velocity of the Bloch-type skyrmionium does rotate 90$^\circ$, parallel to the direction of polarization. These simulation results are in good agreement with our theoretical velocity equations. It should be mentioned that the $y$-component of the N{\'e}el vector $\boldsymbol{n}$ is used to distinguish these two types of skyrmioniums due to the fact that the out-of-plane component is independent of $\Phi(\varphi)$ as mentioned before. Both the internal structure of skyrmioniums and the polarization direction just influence the direction of the velocity, not the magnitude. Therefore, we focus on the dynamics of a N{\'e}el-type skyrmionium driven by the spin current in the following, where $\boldsymbol{p}$ is fixed at $-\boldsymbol{e}_{y}$.

\begin{figure}[t!]
\centerline{\includegraphics[width=0.49\textwidth]{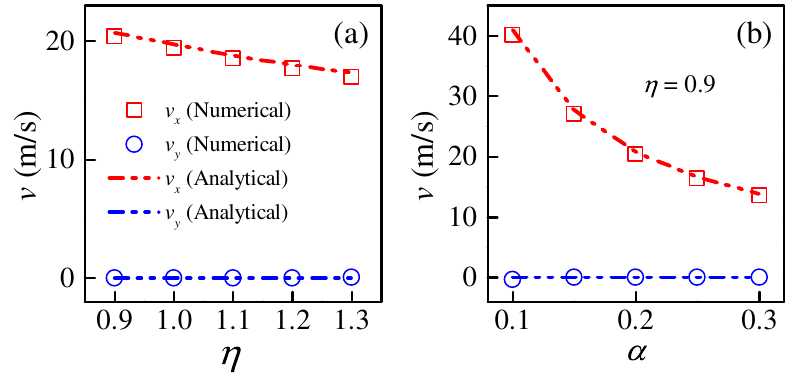}}
\caption{%
Comparison between numerical results and analytical results. 
The ferrimagnetic skyrmionium velocity as a function of (a) the magnetization ratio $\eta=M_{2}/M_{1}$ and (b) the damping coefficient $\alpha$ for $\eta=0.9$. Here, the driving current density $j$ is fixed at 10 MA${\cdot}$cm$^{-2}$ to ensure that the skyrmionium keeps a circle shape without severe distortion. Symbols represent the results from numerical simulations, and the dashed lines indicate the analytical results based on Eq. (~\ref{eq:TvN}).
}
\label{FIG3}
\end{figure}

Figure~\ref{FIG3} illustrates the velocity of a FiM skyrmionium versus the magnetization ratio $\eta=M_{2}/M_{1}$ and the damping coefficient $\alpha$. Note that at the magnetization compensation point, the net magnetic moment vanishes, corresponding to $\eta=M_{2}/M_{1}=1.0$, while the angular momentum compensation point calls for $\eta=M_{2}/M_{1}=1.1$ in this work. From the motion Eqs.~\ref{eq:LLGn} and ~\ref{eq:LLGm}, we find that the spin dynamics strongly depends on $\rho$ and $\sigma$, rather than the net magnetization. Thus, we will discuss more about the angular momentum compensation point in the following sections. In Fig.~\ref{FIG3}, it is seen that the $x$-component of velocity decreases with increasing ratio $\eta$, and is inversely proportional to the damping constant $\alpha$, which are in good agreement with Eq. ~\ref{eq:TvN}.
$v_{y}$ also remains at $0$ as predicted analytically. It should be mentioned that the relation between the FiM skyrmionium velocity and the ratio $\eta$ is monotonous, which is different from that of FiM skyrmions or bimerons~\cite{Woo_NATCOM2018,Caretta_NATTECH2018,Kim_PRB2017}, where the velocity at $\eta=1.1$ reaches maximum.
Such a difference mainly comes from the second term in Eq.~\ref{eq:Thiele}, which is determined by the topological charge $Q$ and the spin density $\sigma$.
For the FiM skyrmions and bimerons with $Q={\pm}1$, $v_{x}$ is maximum and $v_{y}=0$ at the compensation point of the angular momentum ($\eta=1.1$). However, the topological charge $Q=0$ of FiM skyrmioium leads to the $\sigma$-independent velocity (see Eq.~\ref{eq:TvN}) to vary with $\eta$ since the dissipative force is related to $\rho$. Besides, the effect of the parity-breaking exchange constant on the motion of the FiM skyrmionium is also discussed~\cite{SM}.
%
\begin{figure}[t!]
\centerline{\includegraphics[width=0.53\textwidth]{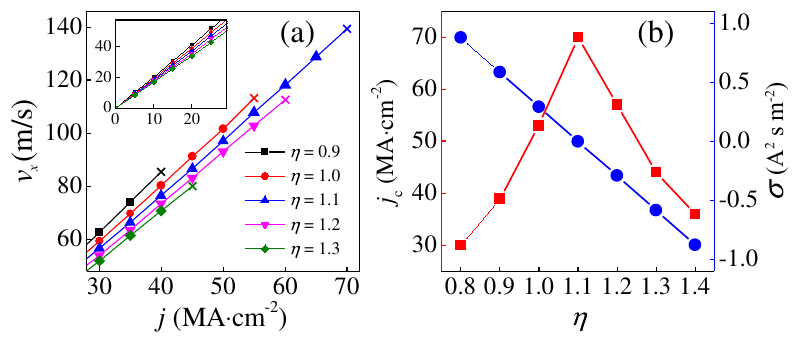}}
\caption{%
Driving current density dependence of the ferrimagnetic skyrmionium velocity. 
(a) Ferrimagnetic skyrmionium velocity as a function of the driving current density $j$ for $\eta=$ 0.9, 1.0, 1.1, 1.2 and 1.3. The inset is zoom-in of the current-velocity relationship when $j<30$ MA${\cdot}$cm$^{-2}$. The cross represents the skyrmionium with severe distortion under such a driving current density.
(b) The critical driving current density that causes a severe distortion for a ferrimagnetic skyrmionium, as a function of the magnetization ratio $\eta$. The blue symbols indicate the spin density $\sigma$ corresponding to different $\eta$.
}
\label{FIG4}
\end{figure}

\begin{figure*}[t]
\centerline{\includegraphics[width=0.9\textwidth]{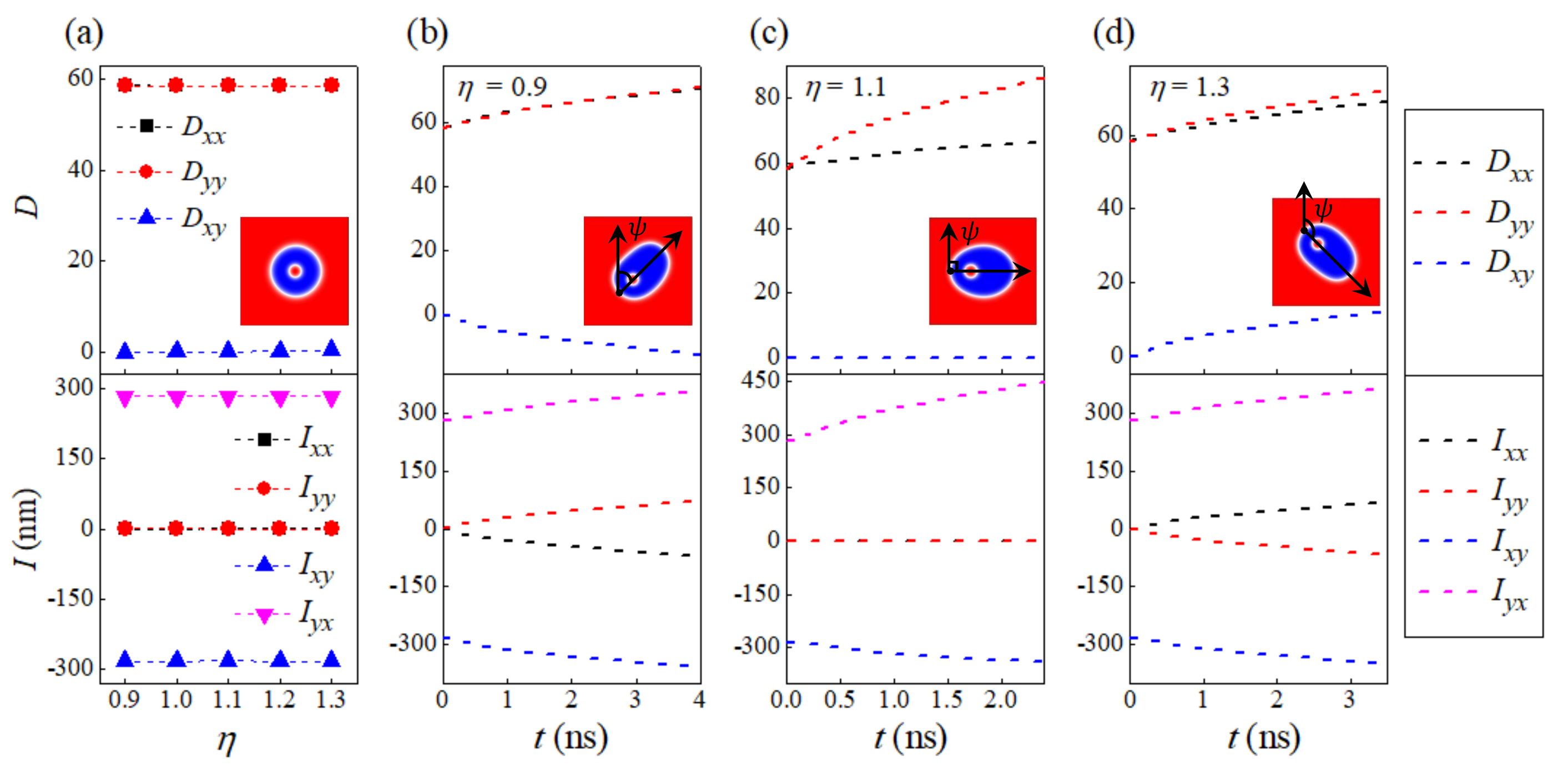}}
\caption{%
Analysis of the skyrmionium deformation.
(a) The components of dissipative tensor $\boldsymbol{\mathcal{D}}$ and tensor $\boldsymbol{\mathcal{I}}$ as functions of the ratio $\eta$ when the skyrmionium moves along the nanotrack without severe distortion. Here, $j=10$ MA$\cdot$cm$^{-2}$.
(b-d) The time evolution of $D_{ij}$ and $I_{ij}$ for $\eta=$ 0.9, 1.1 and 1.3 (corresponding to $\sigma=+0.58, 0.0$ and $-0.58$ A$^2$s$\cdot$m$^{-2}$), where the driving current density is 40, 70 and 45 MA$\cdot$cm$^{-2}$, respectively. Insets show the top view of a ferrimagnetic skyrmionium marked with a deformation angle $\psi$.
}
\label{FIG5}
\end{figure*}

We now study the effect of the driving current density on the dynamics of a FiM skyrmionium. Figure~\ref{FIG4}(a) shows the velocities of a FiM skyrmionium at five different $\eta$ driven by the damping-like SOT.
The FiM skyrmionium shows a similar current-velocity relation at different $\eta$, where the velocity is proportional to the driving current density $j$, corresponding to the Eq.~\ref{eq:TvN}.
However, a large driving force causes severe deformation of the skyrmionium. Here, the level of deformation is distinguished by the aspect ratio of a skyrmionium when it moves to the same specified location. In our work, we measure the aspect ratio of the skyrmionium at 400 nm of a 500-nm-long nanotrack, and the aspect ratio larger than 1.5 is defined as the severe distortion, otherwise it is regarded as a slight deformation.
Fig.~\ref{FIG4}(a) shows that the required current density $j$ to destroy the structure of a FiM skyrmionium is unequal for different $\eta$. For example, the skyrmionium with $\eta=0.9$ is distorted severely at $j=39$ MA${\cdot}$cm$^{-2}$, while the skyrmionium with $\eta=1.2$ shows the same deformation at larger current density $j=58$ MA${\cdot}$cm$^{-2}$.
Figure~\ref{FIG4}(b) illustrates the critical current density $j_{c}$ that forces FiM skyrmionium to be destroyed  seriously versus the ratio $\eta$. The skyrmionium with $\eta=1.1$ is the most robust to resist the deformation, which is attributed to the zero net angular momentum. As shown in Eq.~\ref{eq:v}, the $y$-component of the steady velocity for FiM skyrmions ($Q={\pm}1$) vanishes at the angular momentum compensation point. Namely, there is no skyrmion Hall effect on both the inner and outer skyrmions consisting of the FiM skyrmionium, which is the same as the case of the AFM skyrmionium. In addition, the relationship between the spin density $\sigma$ and the magnetization ratio $\eta$ is also shown in Fig.~\ref{FIG4}(b), where $\sigma$ increases as $\eta$ decreases.

Considering that the tensors $\boldsymbol{\mathcal{D}}$ and $\boldsymbol{\mathcal{I}}$ are determined by the magnetic structure, we extract the components of these tensors to describe the deformation of a FiM skyrmionium.
As shown in Fig.~\ref{FIG5}(a), when the driving current density $j$ is relatively small, the skyrmionium moves along the nanotrack keeping a circular shape without severe distortion.
Therefore, the components $D_{xx}=D_{yy}=58$, $D_{xy}=0$, $I_{xx}=I_{yy}=0$ and $I_{xy}=-I_{yx}=-283$ nm are independent on the ratio $\eta$ and remain unchanged. However, for a large driving current density, the FiM skyrmioniums with different $\eta$ deform in different directions. 
Here, we introduce a deformation angle $\psi$ that is the clockwise angle between the vertical line along the $+y$ direction of the nanotrack and the major axis of the deformed skyrmionium [see Fig.~\ref{FIG5}(b-d)] to parametrize the distortion. It is found that the angle $\psi$ is smaller than 90$^\circ$ when $\eta=0.9$, while $\psi=90^\circ$ for $\eta=1.1$ and $\psi>90^\circ$ for $\eta=1.3$.
The varying deformation originates from the fact that the direction of the Magnus force is determined by the sign of both the topological charge $Q$ and the net spin density $\sigma$.

\begin{figure}[h!]
\centerline{\includegraphics[width=0.39\textwidth]{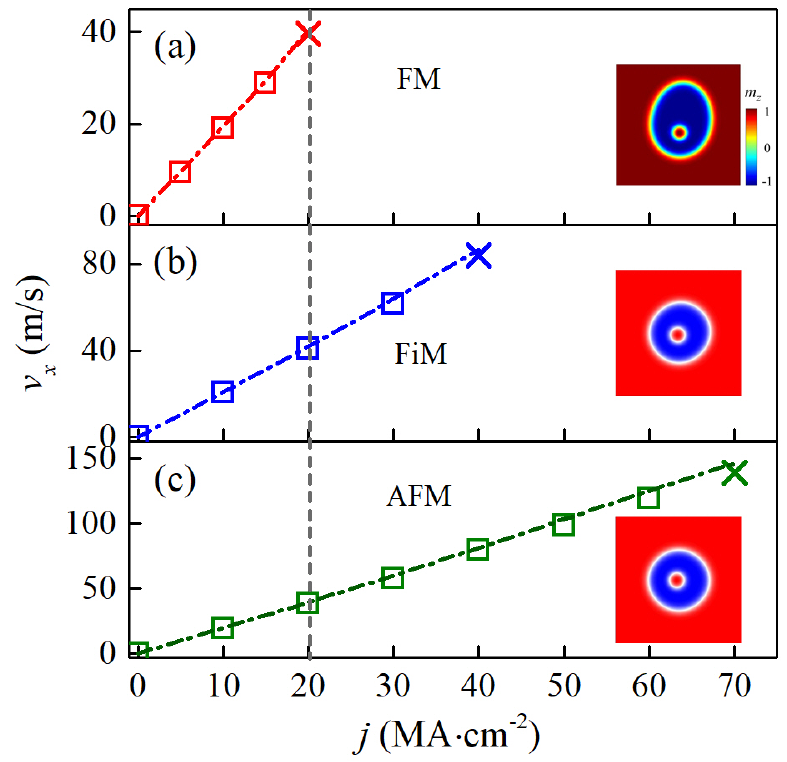}}
\caption{%
Comparison between the current-induced dynamics of (a) FM, (b) FiM ($\eta=0.9$) and (c) AFM skyrmioniums. Insets show the top view of a skyrmionium driven by the selected current density $j=20$ MA$\cdot$cm$^{-2}$ at $t=6$ ns, which is indicated by the vertical dashed line. Symbols represent the results from numerical simulations, and the dashed lines indicate the analytical results based on Eq. (~\ref{eq:TvN}).
}
\label{FIG6}
\end{figure}

\begin{figure*}[t]
\centerline{\includegraphics[width=0.93\textwidth]{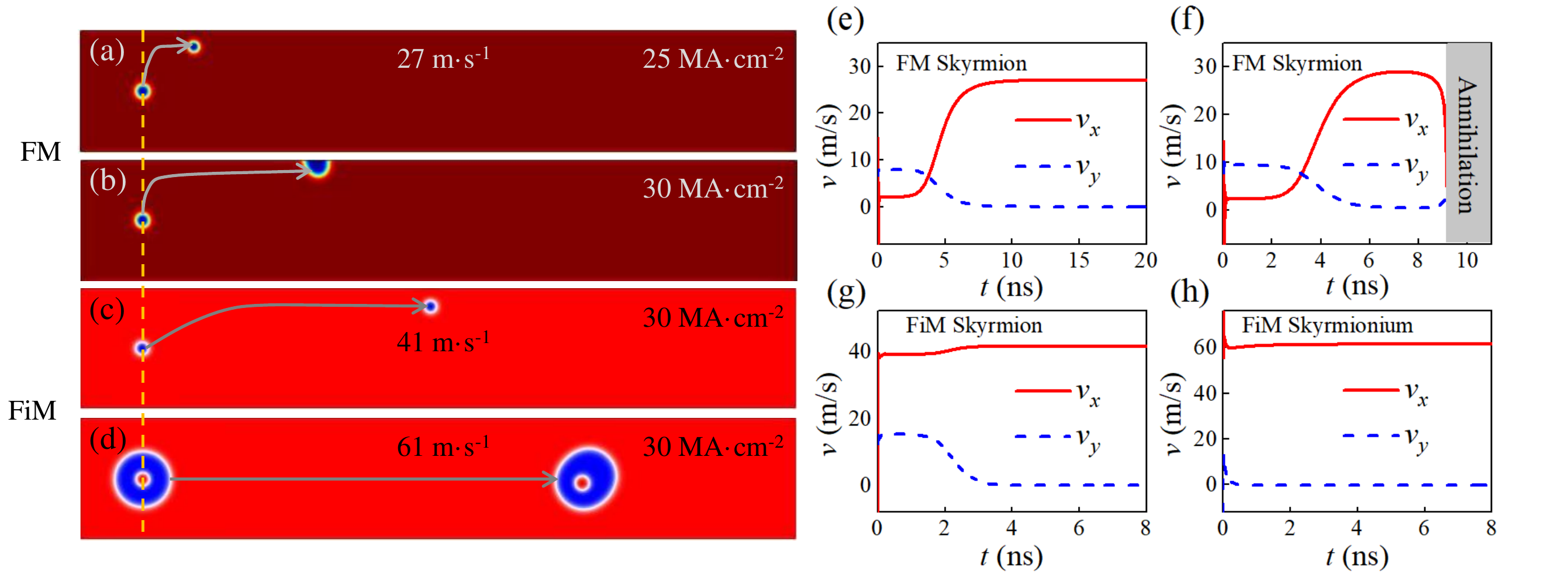}}
\caption{%
Snapshots of (a-b) a FM skyrmion, (c) a FiM skyrmion and (d) a FiM skyrmionium during their motion driven by a spin current, where the yellow dashed line represents the initial position of spin structures, and the gray lines stand for their trajectories.
The time evolution of velocities for these magnetic structures driven by a spin current with (e) $j=25$ MA$\cdot$cm$^{-2}$ and (f-h) $j=30$ MA$\cdot$cm$^{-2}$.
Here, $\eta=0.9$ for a typical uncompensated FiM system.  
}
\label{FIG7}
\end{figure*}

As shown in Fig.~\ref{FIG5}(b-d), $\eta=0.9$ ($\sigma=+0.58$ A$^2$s$\cdot$m$^{-2}$), the drift speed $v_{y}$ of the inner skyrmion is along the $-y$ direction, while the Magnus force acting on the outer skyrmion points up. Besides, the magnitudes of $v_{x}$ and $v_{y}$ are proportional to the size of skyrmions. As a consequence, $\psi<90^\circ$, and the components of tensors $\boldsymbol{\mathcal{D}}$ as well as $\boldsymbol{\mathcal{I}}$ obviously vary with time. On the contrary, when $\eta=1.3$ ($\sigma=-0.58$ A$^2$s$\cdot$m$^{-2}$), the drift speed $v_{y}$ of the inner skyrmion and the outer skyrmion are upward and downward respectively, which induce a deformation angle $\psi$ larger than 90$^\circ$. As discussed above, the skyrmions have no drift speed along the $y$-axis at the compensation point of the angular momentum, so that the two skyrmions consisting of the FiM skyrmionium move along the $x$-axis at different velocities leading to $\psi=90^\circ$. It is worth mentioning that, when $\eta=1.1$ ($\sigma=0.0$ A$^2$s$\cdot$m$^{-2}$), the components $D_{xy}$, $I_{xx}$ and $I_{yy}$ are still $0$ and do not vary with time. However, when $\eta {\neq} 1.1$, these quantities are nonzero, indicting the different deformations. In general, $D_{xy}<0$ ($I_{xx}<0$ and $I_{yy}>0$) corresponds to the deformation angle $\psi<90^\circ$, while $D_{xy}>0$ ($I_{xx}>0$ and $I_{yy}<0$) is related to the deformation angle $\psi>90^\circ$. The final configurations of these deformed FiM skyrmioniums at such current densities are provided in Ref.~\cite{SM}.

We now compare the current-induced dynamics of FM, FiM and AFM skyrmioniums.
In Fig.~\ref{FIG6}, we demonstrate that the response of a skyrmionium to the driving current in the three magnetic systems. It is seen that a relatively small driving current density results in the distortion of a FM skyrmionium during its motion, while the AFM skyrmionium is the most difficult to be destroyed. The FiM skyrmionium is the intermediate one between the cases of FM and AFM skyrmioniums.
Moreover, driven by the same current density of $j=20$ MA${\cdot}$cm$^{-2}$ and compared with FiM and AFM skyrmioniums, the FM skyrmionium is most fragile. Taking the micro-structures of the three systems into account, the net angular momentum is completely canceled in the AFM system due to the two sublattices coupled antiferromagnetically with the same magnetic moments and the same gyromagnetic ratio.
Different from the AFM system, there is a resultant angular momentum in the FiM system, where the two sublattices normally have different spin densities $M_{i}/\gamma_{i}$.
On the other hand, the net angular momentum of the FiM system is still smaller than that of the FM system.
Note that the Magnus force depends on $\sigma$ as shown in Eq.~\ref{eq:Thiele}.
From the point of view of applications, the FiM skyrmionium may be a good choice since it not only suppresses the skyrmion Hall effect more effectively than the FM skyrmionium, but also is easier to be detected compared with the AFM one.

Here, the current-induced dynamics of a FiM skyrmionium is also compared with that of a FM skyrmion and a FiM skyrmion. During the motion driven by a spin current, the FM skyrmion shows an inevitable skyrmion Hall effect due to the Magnus force associated with the nonzero topological charge. In a confined nanotrack, as shown in Fig.~\ref{FIG7}(a), this Magnus force is canceled by the repulsive force arisen from the boundary so that the skyrmion moves along the edge with a steady speed. However, for a large driving current density, the energy barrier of the edge is insufficient to confine a skyrmion in a nanotrack. Consequently, the skyrmion disappears at the edge [see Fig.~\ref{FIG7}(b)]. Considering that the net angular momentum of two sublattices are not completely cancelled, the FiM skyrmion shows the similar motion behaviors. From Figs.~\ref{FIG7}(c) and (d), it is seen that the steady speed of a FiM skyrmion is smaller than that of a FiM skyrmionium driven by the same current density. Moreover, the FiM skyrmionium moves along the centerline of the nanotrack without drift speed as discussed before.
Figures~\ref{FIG7}(e-h) also show the time evolution of velocities for these spin configurations during their motion. For the undamaged FM skyrmion and FiM skyrmion, the $x$-component of the velocity begins with a constant, and then increases as the skyrmion approaches the edge and eventually reaches a new steady value, while the $y$-component decreases to zero. However, the velocity of a skyrmionium remains unchanged with a relatively large value. Therefore, compared with skyrmions, the FiM skyrmionium as a carrier of information can effectively prevent the accumulation and annihilation of skyrmions at the edge, and potentially improve the access speed of storage devices.

\section{Conclusion}
\label{se:Conclusion}

In conclusion, we have analytically and numerically investigated the current-induced dynamics of a FiM skyrmionium in a nanotrack.
Our results show that, at the angular momentum compensation point, the FiM skyrmionium is most robust to resist the deformation due to the zero intrinsic skyrmion Hall effect, which is same as the case of an AFM skyrmionium.
Nevertheless, the position of a FiM skyrmioniums is observable due to a nonzero magnetization.
It is found that the direction of distortion depends on the sign of the net angular momentum.
Above the angular momentum compensation point, the deformation angle is smaller than 90$^\circ$.
However, the deformation angle is larger than 90$^\circ$ below the angular momentum compensation point.
We have also demonstrated the change in components of two tensors $\boldsymbol{\mathcal{D}}$ and $\boldsymbol{\mathcal{I}}$ for a FiM skyrmionium during the motion to describe its deformation.
Furthermore, we have made a comparison between the current-induced dynamics of FM, FiM and AFM skyrmioniums. The motion of a FiM skyrmionium is also compared with that of FM and FiM skyrmions.
Our results open a new filed of the skyrmionium physics in the FiM system and could provide guidelines for the design of future spintronic devices based on ferrimagnets.

\begin{acknowledgments}

This research was supported by Guangdong Special Support Project (2019BT02X030), Shenzhen Fundamental Research Fund (Grant No. JCYJ20210324120213037), Shenzhen Peacock Group Plan (KQTD20180413181702403), Pearl River Recruitment Program of Talents (2017GC010293) and National Natural Science Foundation of China (11974298, 61961136006).
X.L.'s PhD study was financially supported by the National Natural Science Foundation of China (Grant No. 12004320).
X.Z. was an International Research Fellow of the Japan Society for the Promotion of Science (JSPS). X.Z. was supported by JSPS KAKENHI (Grant No. JP20F20363).
J.X. acknowledges the support by the National Natural Science Foundation of China (Grant No. 12104327).
M.E. acknowledges the support by the Grants-in-Aid for Scientific Research from JSPS KAKENHI (Grant Nos. JP17K05490 and JP18H03676) and the support by CREST, JST (Grant Nos. JPMJCR16F1 and JPMJCR20T2).
X.X.L. acknowledges the support by the Grants-in-Aid for Scientific Research from JSPS KAKENHI (Grant Nos. JP20F20363 and JP21H01364).
\end{acknowledgments}

\begin{appendix}

\section{ANALYTICAL DERIVATION OF THE MOTION EQUATIONS}
\label{se:Appendix A}

To derive the motion equation of magnetization in a ferrimagnetic system, we start from the Laudau-Lifshitz-Gilbert (LLG) equations with the spin-orbit torque for the two sublattices:
\begin{align}    
\dot{\s}_1\!=\!&-\gamma_{1}\s_1\!\times\!\He_1+\alpha\s_1\!\times\!\dot{\s}_1+\gamma_{1}B_{D1}\s_1\!\times\!(\p\!\times\!\s_1), \label{eq:LLGs1}  \\
\dot{\s}_2\!=\!&-\gamma_{2}\s_2\!\times\!\He_2+\alpha\s_2\!\times\!\dot{\s}_2+\gamma_{2}B_{D2}\s_2\!\times\!(\p\!\times\!\s_2), \label{eq:LLGs2} 
\end{align}
where $\He_i=-\delta{\varepsilon}⁄(\mu_0 M_i \delta\s_i)$ and  $B_{Di}=(\mu_B \theta_{SH}j)⁄(\gamma_i eM_i t_z)$ are the effective fields associated with various energies in the system and the damping-like torque induced by the spin current, respectively. Multiplying the Eqs. (\ref{eq:LLGs1}) and (\ref{eq:LLGs2}) by $M_1/\gamma_1$ and $M_2/\gamma_2$, respectively, and then finding the addition and the subtraction of these two equations, we obtain the following coupled equations of motion:
\begin{eqnarray}    
\label{eq:LLGmn0}
\rho\dot{\n}&+&\sigma\dot{\m}=-(\m\times\f_n+\n\times\f_m)  \nonumber \\ 
&+&\alpha[\rho (\n\times\dot{\m}+\m\times\dot{\n})+\sigma (\m\times\dot{\m}+\n\times\dot{\n})]  \nonumber \\ 
&+&u_1[\m\times(\p\times\n)+\n\times(\p\times\m)]  \nonumber \\ 
&+&u_2[\n\times(\p\times\n)+\m\times(\p\times\m)],  \\ 
\rho\dot{\m}&+&\sigma\dot{\n}=-(\m\times\f_m+\n\times\f_n)  \nonumber \\ 
&+&\alpha[\rho (\m\times\dot{\m}+\n\times\dot{\n})+\sigma (\n\times\dot{\m}+\m\times\dot{\n})]  \nonumber \\ 
&+&u_1[\n\times(\p\times\n)+\m\times(\p\times\m)]  \nonumber \\ 
&+&u_2[\m\times(\p\times\n)+\n\times(\p\times\m)].  
\end{eqnarray}
Here, $\rho=M_1/\gamma_1+M_2/\gamma_2$, $\sigma=M_1/\gamma_1 -M_2/\gamma_2$ , $\f_m=\f_1+\f_2$, $\f_n=\f_1-\f_2$ and $u_1=\beta_1+\beta_2$, $u_2=\beta_1-\beta_2$ with $\beta_i=B_{Di} M_i$ and $\f_i=\He_i M_i$. It is important to noted that the rule for the linear combination of two variables is used in this derivation, $Ax+By=(1/2) [(A+B)(x+y)+(A-B)(x-y)]$ and $Ax-By=(1/2) [(A+B)(x-y)+(A-B)(x+y)]$. Considering the fact that $|\m|\ll|\n|\approx 1$ for the colinear ferrimagnets, and keeping leading-order terms, the above equations are reduced as
\begin{eqnarray}    
\rho\dot{\n}=\f_m&\times&\n+\alpha(\rho\n\times\dot{\m}+\sigma \n\times\dot{\n}) \nonumber \\ 
&+&u_2\n\times(\p\times\n)+\T_{\text{nl}}^n,  \label{eq:LLGnA} \\ 
\rho\dot{\m}+\sigma\dot{\n}=&\f&_m\times\m+\f_n\times\n+\alpha\rho\n\times\dot{\n} \nonumber \\ 
&+&u_1\n\times(\p\times\n)+\T_{\text{nl}}^m,  \label{eq:LLGmA} 
\end{eqnarray}
where, $\T_{\text{nl}}^n=u_1 \n\times(\p\times\m)$ and $\T_{\text{nl}}^m=\alpha\sigma\n\times\dot{\m}+u_2 \n\times(\p\times\m)$ are the weak nonlinear terms that will be discarded in the following derivation. The micromagnetic simulation of the spin dynamics in this work is based on numerically solving Eqs. (\ref{eq:LLGnA}) and (\ref{eq:LLGmA}).

Substituting $\f_m=-(1/\mu_0)  [\lambda\m+L(\partial_x \n+\partial_y \n)]$ into Eq. (\ref{eq:LLGnA}), we obtain the total magnetization $\m$ that depends on the spatial N{\'e}el vector $\n$.
\begin{eqnarray}    
\label{eq:m}
\!\!\!\!\!\!\!\m&=&\frac{\mu_0}{\lambda}\{-\rho(1+\alpha^2 )\n\times\dot{\n}+\alpha[\n\times\f_n-u_1\n\times(\p\times\n)] \nonumber \\
&\;&-u_2 \p\times\n\}-\frac{L}{\lambda}(\partial_x \n+\partial_y \n), 
\end{eqnarray}
where the dissipation term $\alpha[\n\times\f_n-u_1 \n\times(\p\times\n)]$ can be ignored. Rewriting the effective field $\f_n=\f_n^{*}+(A^{*}/\mu_0) (\partial_{xx}+\partial_{yy}+2\partial_{xy})\n+(L/\mu_0) (\partial_x\m+\partial_y\m)$ with $A^{*}=A/2$ and substituting $\m$ into $\f_n\times\n$, we obtain
\begin{eqnarray}    
\label{eq:fnxn}
\f_n\times\n=&\f&_n^{*}\times\n+\frac{A}{2\mu_0}(\partial_{xx}+\partial_{yy}+2\partial_{xy})\n\times\n+\frac{L}{\mu_0} \nonumber  \\
\langle\frac{\mu_0}{\lambda}\{-\rho&(&1+\alpha^2)\n\times(\partial_x+\partial_y)\dot{\n}-u_2\p\times(\partial_x+\partial_y)\n\} \nonumber  \\
&-&\frac{L}{\lambda}(\partial_{xx}\n+\partial_{yy}\n+2\partial_{xy}\n)\rangle\times\n. 
\end{eqnarray}
Since $\lambda=4A⁄a^2$  and $L=\sqrt{2}A⁄a$, we deduce that $A/2=L^2/\lambda$, and then the second and the last terms on the right side of the Eq.(\ref{eq:fnxn}) can be cancelled. We substitute $\m$ and $\f_n\times\n$ into Eq. (\ref{eq:LLGmA}),
\begin{eqnarray}
\label{eq:m+fnxn}
\frac{\mu_0\rho}{\lambda}\{&-&\rho(1+\alpha^2)\n\times\ddot{\n}-u_2 \p\times\dot{\n}\}-\frac{L\rho}{\lambda}(\partial_x \dot{\n} +\partial_y\dot{\n})  \nonumber  \\
=&-&\sigma\dot{\n} +\f_m\times\m+\f_n^{*}\times\n+\frac{L}{\lambda}\{-\rho(1+\alpha^2 )  \nonumber  \\
&\n&\times (\partial_x+\partial_y )\dot{\n}- u_2 \p\times(\partial_x+\partial_y)\n\}\times\n  \nonumber \\
&+&\alpha\rho\n\times\dot{\n}+ u_1 \n\times(\p\times\n). 
\end{eqnarray}
Supposing that $(1+\alpha^2 )\approx 1$ and multiplying the Eq. (\ref{eq:m+fnxn}) by $\n$, we obtain the following closed equation of the N{\'e}el vector $\n$
\begin{eqnarray}    
\label{eq:closedA}
\frac{\mu_0 \rho^2}{\lambda}&\n&\times(\n\times\ddot{\n})=\sigma\n\times\dot{\n}+\alpha\rho\dot{\n}+u_1 \p\times\n \nonumber \\
&-&\frac{\mu_{0}\rho{u_2}}{\lambda}\boldsymbol{n}\times(\p\times\dot{\boldsymbol{n}})-\boldsymbol{n}\times(\boldsymbol{f}^\ast_{n}\times\boldsymbol{n}) \nonumber  \\
&+&\frac{Lu_{2}}{\lambda}\boldsymbol{n}\times\{[\boldsymbol{p}\times(\partial_{x}+\partial_{y})\boldsymbol{n}]\times\boldsymbol{n}\}.  
\end{eqnarray}
For a centrosymmetric magnetic soliton, the order parameter $\n$ is described by both the position $r$ and the azimuthal angle $\varphi$, i.e., $\n(r,\varphi)=[\sin\theta(r)\cos\Phi(\varphi),\sin\theta(r)\sin\Phi(\varphi),\cos\theta(r)]$. Taking the scalar product of Eq. (\ref{eq:closedA}) with $\partial_i\n$, and integrating over the space, we obtain the motion equation 
\begin{equation}    
\label{eq:ThieleA}
M\ddot{\R}+\sigma\G\times\dot{\R}+\alpha\rho\D\dot{\R}-u_1\I\p=\0,  
\end{equation}
where $M=\mu_0 \rho^2d⁄\lambda$ is the effective mass of the soliton, $\G=(0,0,G)$ is the topology-dependent gyrovector, $\D$ and $\I$ are tensors related to the damping term and the SOT, respectively. Here, 
\begin{align}    
\label{eq:DGI}
d_{ij}=\int{}(\partial_i \n\cdot\partial_j \n)dxdy=&\delta_{ij}d  \nonumber  \\
=\delta_{ij}\pi\int{}(\theta_r^2+\sin^2 \theta/r^2)r&dr,  \\
G=\int{}[\n\cdot(\partial_x\n\times\partial_y \n)]dxdy=2\pi&\int{}\sin\theta\theta_r dr, \\
I_{ij}=\int{}(\n\times\partial_i \n)_jdxdy&, 
\end{align}

\begin{align}
\int{}\partial_i\n\cdot[\n\times(\p\times\dot{\n})]dxdy=&0,  \\
\int{}\partial _i\n\cdot [\n\times(\f_n^{*}\times\n)]dxdy=&0,  \\
\int{}\partial_i \n\cdot  \{\n \times[\p\times(\partial_x+\partial_y )\n]\times\n\}&dxdy=0. 
\end{align}
This term $(\n\times\partial_i \n)_j$ denotes the $j$-component of the vector $(\n\times\partial_i \n)$. Different from the gyrovector $\G$ and the tensor $\D$, the tensor $\I$ strongly depends on the angle $\Phi(\varphi)$ due to the fact that 
\begin{align}   
\label{eq:Ixy}
&(\n\times\partial_x \n)_x=-\sin\Phi\cos\varphi\theta_r+\sin\theta\cos\theta\frac{\sin\varphi\cos\Phi}{r}, \\
&(\n\times\partial_x \n)_y=\cos\Phi\cos\varphi\theta_r+\sin\theta\cos\theta \frac{\sin\varphi\sin\Phi}{r},  \\
&(\n\times\partial_y \n)_x=-\sin\Phi\sin\varphi\theta_r-\sin\theta\cos\theta\frac{\cos\varphi\cos\Phi}{r},\\
&(\n\times\partial_y \n)_y=\cos\Phi\sin\varphi\theta_r-\sin\theta\cos\theta \frac{\cos\varphi\sin\Phi}{r}. 
\end{align}
In general, the steady velocity is written as
\begin{equation}
\label{eq:vA} 
\begin{bmatrix}
v_{x}\\
v_{y}
\end{bmatrix}
=\frac{u_{1}}{{\alpha}^2{\rho}^2d^2+{\sigma}^2G^2}
\begin{bmatrix}
{\alpha}{\rho}d & {\sigma}G \\
{-\sigma}G & {\alpha}{\rho}d
\end{bmatrix}\begin{bmatrix}
I_{xx} & I_{xy} \\
I_{yx} & I_{yy}
\end{bmatrix}\begin{bmatrix}
p_{x}\\
p_{y}
\end{bmatrix}. 
\end{equation}
For a N{\'e}el-type skyrmionium, $\Phi=\varphi$, $I_{xy}=-I_{yx}=I=\pi\int{(r\theta_r+\sin\theta\cos\theta)dr}$ and $I_{xx}=I_{yy}=0$, above velocity equation becomes 
\begin{equation}
\label{eq:TvNA} 
\left[
\begin{array}{c}
v_{x}\\
v_{y}
\end{array}
\right]_{\text{N{\'e}el}}=\frac{u_{1}I}{\alpha\rho d}\left[
\begin{array}{c}
p_{y}\\
-p_{x}
\end{array}
\right]. 
\end{equation}
Similarly, for the Bloch-type skyrmionium, $\Phi=\varphi+\pi⁄2$, $I_{xx}=I_{yy}=-I$ and $I_{xy}=I_{yx}=0$, the velocity is given by
\begin{equation}
\label{eq:TvBA} 
\left[
\begin{array}{c}
v_{x}\\
v_{y}
\end{array}
\right]_{\text{Bloch}}=\frac{-u_{1}I}{\alpha\rho d}\left[
\begin{array}{c}
p_{x}\\
p_{y}
\end{array}
\right]. 
\end{equation}
Therefore, one can find that the velocity of a skyrmionium depends on both the internal spin distribution and the polarization direction of the spin current.

\end{appendix}



\end{document}